\documentclass[11pt, twoside, epsf]{article}

\usepackage{asp2004}
\usepackage{epsf}
\usepackage{psfig}
\usepackage{lscape}
\usepackage{graphicx}

\begin{document}
\title{Extreme Scattering Events: insights into the interstellar medium on AU-scales}   
\author{Mark A. Walker}   
\affil{MAW Technology Pty Ltd, 3/22 Cliff Street, Manly 2095, Australia}    

\begin{abstract}
Several radio-wave scintillation phenomena exhibit properties which are difficult to accommodate within the standard propagation model based on distributed Kolmogorov turbulence in the ionised ISM; here we discuss one such phenomenon, namely Extreme Scattering Events. By analysis of the data we demonstrate that these events are caused by ionised gas associated with self-gravitating, AU-sized gas clouds. The data also show that the ionised gas is confined by ram pressure, with the clouds moving at hundreds of km/s relative to the diffuse ISM and causing strong shocks. These conclusions are supported by a quantitative model in which heat from the shocked ISM evaporates gas from the surface of a cold cloud; this model readily explains the physical conditions which are required for Extreme Scattering and yields passable reproductions of the light-curves. The magnetotail of the cloud provides a site in which two other ``anomalous'' radio-wave propagation phenomena -- IntraDay Variability of quasars, and pulsar parabolic arcs -- can plausibly  arise, thus linking three anomalous propagation phenomena in a single physical model.  Locally there must be thousands of these neutral clouds per cubic parsec and by mass they are the primary constituent  of interstellar matter.
\end{abstract}


\section{Introduction}
Extreme Scattering Events (ESEs) were discovered by \citet{F87} in a radio flux monitoring program covering a large number of bright quasars over a period of several years. They are propagation effects in which the radio flux is strongly influenced by refraction in ionised gas in our own Galaxy \citep*{F87, RBC87, F94}.  Here we shall concentrate mainly on the single most spectacular ESE, which was recorded in the 2.7 and 8.1~GHz fluxes of Q0954+658 \citep{F87}; this event is shown in figure 1.  As noted by \citet{RBC87} \citep[see also][]{W01} this behaviour  is lens-like, implying an enhancement $\sim10^{16}\,{\rm cm^{-2}}$ in the electron column density  on transverse scales $\sim10^{13}\,{\rm cm}$.  A key question is whether the electron column-density enhancement is due to a volume-density enhancement with a line-of-sight depth comparable to the observed transverse scale? If so the implied pressure of the ionised gas is $\sim 10^{7}\,{\rm K\,cm^{-3}}$, at least three orders of magnitude above typical ISM pressures, implying a short lifetime and making it difficult to understand how the over-density arose in first place.  Spreading the column-density enhancement uniformly over a much greater extent along the line-of-sight alleviates these problems. However, the existing data exclude this possibility, as discussed in the following section, forcing us to accept the existence of localised high pressure regions in the ISM.

\begin{figure}[!ht]
\begin{center}
\includegraphics[width = 12cm, clip = true, bb = 70 350 550 700]{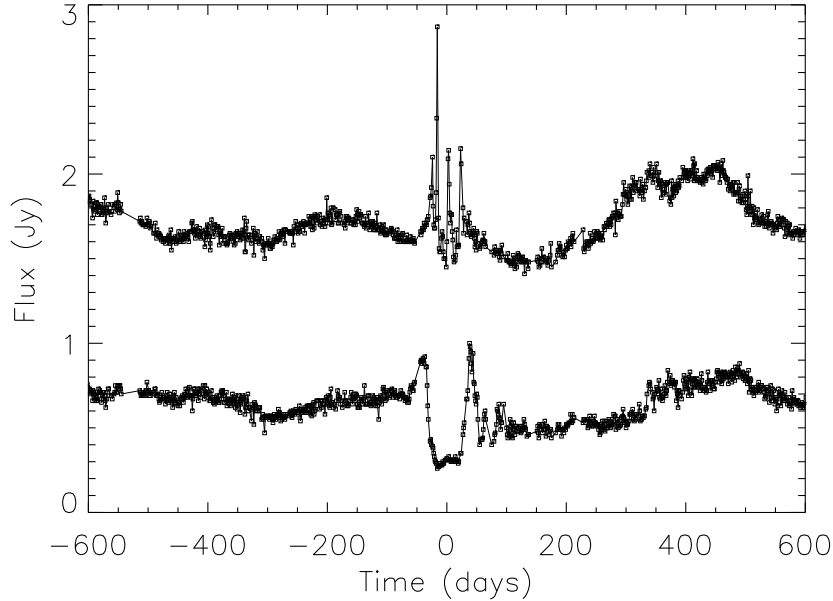}
\end{center}
\vskip -1cm
\caption{The Extreme Scattering Event in  Q0954+658, at 8.1~GHz (top curve) and 2.7~GHz (lower curve), from \citet{F87}. An offset of $+1$~Jy has been added to the high-frequency data for clarity of presentation. The mid-point (symmetry point) of the event at ${\rm Time = 0}$ corresponds to MJD44650.00.  These data  taken from {\tt http://ese.nrl.navy.mil/.\/} }\label{fig1}
\end{figure}

\section{Symmetry and its implications}
It is a remarkable fact that the radio light-curves of ESEs are roughly time-symmetric \citep{W01}; symmetry is not something that we expect of interstellar gas.   The symmetry is only approximate -- there are substantial differences between the two halves of each light-curve in figure 1 -- but deviations from symmetry exist in any real gas distribution even if they are not present in our idealised models, so in the first instance we need only attend to the symmetry. What kind of symmetry is it? The rough temporal symmetry evident in the light-curves tells us that there is a reflection symmetry in the electron column-density distribution along the trajectory traced out by the line-of-sight to the quasar.  If this symmetry is more than a mere fluke then it should hold for all possible straight-line trajectories across the lens plane. Thus the lens geometries are restricted to either (i) mirror-symmetric lenses with translational invariance parallel to the symmetry axis,  or (ii) axisymmetric lenses.  These two possibilities are quite distinct in the light curves they generate: a mirror-symmetric lens always manifests flux conservation in the light-curve (i.e. the flux averaged over the event is the same as the unlensed flux), whereas an axisymmetric lens in general does not. This distinction is independent of the particular electron column-density profiles, being a direct consequence of the differences between the generic forms of the Jacobian of the lens mapping in these two cases. 

\begin{figure}[!ht]
\begin{center}
\includegraphics[width = 12cm, clip = true, bb = 70 350 550 700]{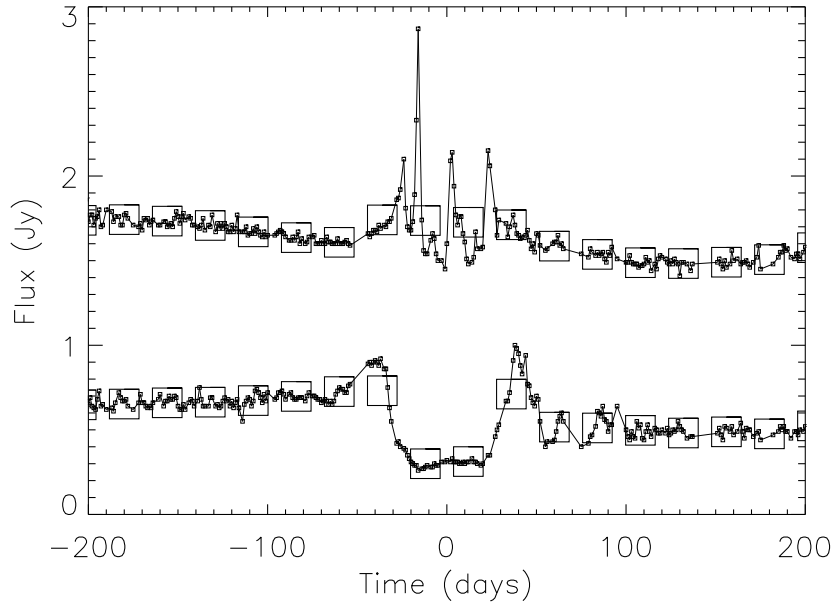}
\end{center}
\vskip -1cm
\caption{As figure 1, but showing only 400~days around the event. Overlaid on the original data are 24-day average fluxes, plotted as squares;  the time intervals for averaging are such that ${\rm Time=0}$ corresponds to an interval boundary. Neither of the two frequencies manifests flux conservation in the light-curve, as is evident from the high values of the four central squares in the 8.1~GHz (upper) data and the low values of the central two squares in the 2.7~GHz (lower) data.}\label{fig2}
\end{figure}

Figure 2 shows the 2.7 and 8.1~GHz fluxes of Q0954+658 with each point averaged over an interval of 24 days. For both frequencies the averaging intervals are measured from the apparent symmetry point (${\rm Time=0}$), so to the extent that the event is time-symmetric the two halves should display the same behaviour.  Within the event itself -- which we take to be the central four averaged data points at 8.1~GHz and the central six averaged data points at 2.7~GHz -- this is almost satisfied. The one discrepancy is the outermost pair of the six points at 2.7~GHz. Both of these points lie close to the baseline flux value but after the event the baseline is lower than that beforehand; we put aside the interpretation of this baseline shift until \S3.  For now the reader should ignore the right-hand (${\rm Time > 0}$) portion of the low-frequency light-curve, bearing in mind that we are attempting to discriminate between two different symmetric models so asymmetries are a distraction. The left-hand portion of the 2.7~GHz light-curve shows a mean flux during the event which is lower than the unlensed flux, while at 8.1~GHz both halves of the data clearly show an average lensed flux which is {\it higher\/} than the unlensed flux. 

One could argue that at 2.7~GHz there might be some flux refracted through large angles that we are missing in our accounting by restricting attention to the central data points. This is a contrived argument because the 2.7~GHz flux drops rapidly away from the peaks at ${\rm |Time|\simeq40\;days}$, suggesting that we are not missing any significant contribution in our averaging. Moreover this argument fails to explain the behaviour at 8.1~GHz, where the refraction angles are much smaller and the average lensed flux is {\it higher\/} than the unlensed flux. For an axisymmetric lens the average flux during a lensing event may be greater than or less than the unlensed flux, depending on the particular column-density profile, the trajectory of the lens relative to the source, and the radio frequency. We conclude that these data are only consistent with a lens which has an axisymmetric distribution of ionised gas. 

An axisymmetric column-density distribution can arise in two ways: either the ionised gas has a spherically symmetric distribution, or else the three dimensional distribution has cylindrical symmetry and the axis of symmetry is aligned with the line-of-sight.  In the latter circumstance we would expect there to be many misaligned cylinders with high electron column-density for every one which is aligned, implying many non-symmetric ESEs with comparable depth of modulation for every symmetric one. That is not what we see. Of the nine radio-sources presented in \citet{F94}, the two ESEs with the largest fractional variations are those in Q0954+658 and Q1749+096, both of which are roughly time-symmetric events. In so far as the lower quality data for the Q1749+096 event admit a comparison its light-curves bear a substantial similarity to those of  Q0954+658, suggesting a common physical origin. So we conclude that the lenses appear axisymmetric because they have an underlying spherical symmetry; this indicates that they are associated with self-gravitating gas.

\section{Asymmetry and its implications}
A striking asymmetry of the Q0954+658 light-curves seen in figure 2 is that egress of the event shows substantial levels of rapid variability, whereas no such variability is evident at ingress; and this is true at both high and low radio frequencies. This tells us that there is much more small-scale structure in the ionised gas on the trailing side of the lens than on the leading side. Presumably this difference is also the reason why the 2.7~GHz average fluxes are significantly lower immediately after the refractive event than in the corresponding period immediately before it --- a substantial fraction of the source flux has been scattered through angles which are larger than the apparent size of the scattering region.  (In this interpretation {\it both\/} halves of the 2.7~GHz light-curve have an average lensed flux which is less than the unlensed flux.) In other astrophysical contexts small scale density structure in ionised gas is the norm, as transonic velocity differences provide a source of free energy from which flow instabilities tend to develop. The low level of small-scale structure on the leading edge of the lens therefore suggests that the flow speed in the high-density ionised gas is highly subsonic there, implying an external pressure contribution on one side which is comparable to the thermal pressure in the lens. If the evolution of the lensing geometry during the event is due primarily to motion of the lens, rather than the observer, then this side is in the forward hemisphere of the cloud and it is straightforward to identify the source of the pressure asymmetry as the ram pressure of the diffuse interstellar medium, seen in the frame of the lens.

If the  cloud moves at speed $v$ relative to the diffuse ISM, which has sound-speed $c_s$, then the ram pressure exerted by the ISM is greater than its thermal pressure by a factor $(v/c_s)^2$.  This offers a simple explanation for the very high pressure environment implied by the results of \S2: we require only that $v\sim30\,c_s$ to explain pressures which are $\sim1000\times$~ambient, so a cloud moving at hundreds of km/s in the cold or warm ISM (but not the hot ISM) suffices. Under these circumstances a strong shock forms in the diffuse ISM upstream of the cloud. The shocked gas is hot ($\sim10^6\;{\rm K}$), whereas the cloud is cold, and a temperature gradient is thus established with inward heat flux evaporating material from the surface of the cloud. This evaporation flow is roughly isobaric and the greatest radio-wave refraction occurs in the coolest ionised gas, close to the cloud surface.

\section{Quantitative lens  model}
To describe the evaporation flow quantitatively we model the shocked gas by a static medium of the same temperature and pressure as would result from an infinite plane-parallel shock travelling at the same speed and with the same upstream conditions. A cold, spherical cloud is immersed in this medium and we seek a steady-state solution in which heat from the shocked ISM evaporates gas from the cloud surface. Although this is a simplified model it has the merit that it allows us to make use of published analytic results: in this section we rely on the analysis of \citet{CM77}, who derived general results for conductive evaporation flows.

\begin{figure}[!ht]
\begin{center}
\includegraphics[width = 12cm, clip = true, bb = 70 350 550 700]{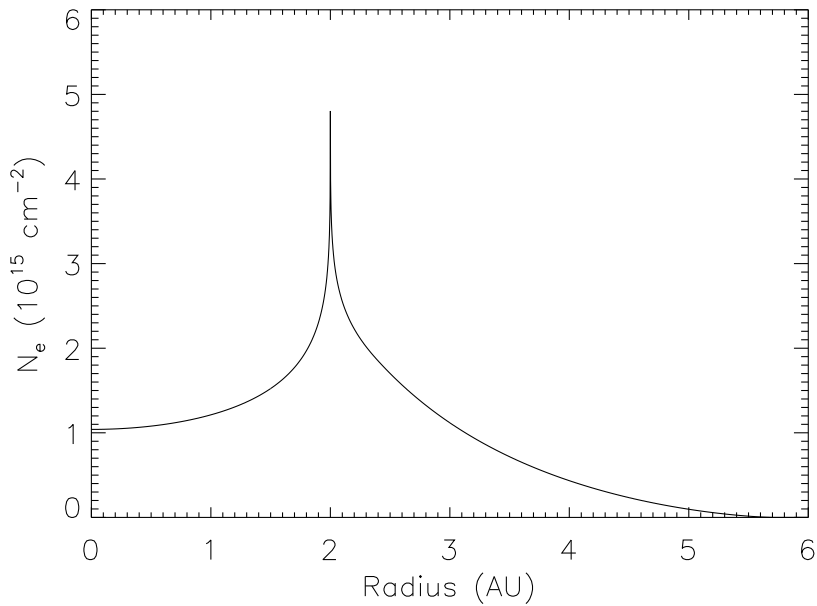}
\end{center}
\vskip -1cm
\caption{The electron column density excess due to ionised gas evaporated from the surface of a cold cloud immersed in hot gas -- as described in \S4.}\label{fig3}
\end{figure}

We adopt temperature and pressure of $1000\;{\rm K}$ and $3000\;{\rm K\,cm^{-3}}$ for the upstream ISM. With a shock speed of $300\;{\rm km\,s^{-1}}$ the post-shock density and temperature are $12\;{\rm cm^{-3}}$ and $10^6\;{\rm K}$, respectively. The cloud radius is assumed to be 2~AU.  The resulting model for the excess electron column-density profile in the evaporation flow is as shown in figure 3. To arrive at this result we have employed the temperature and pressure profiles of \citet{CM77}, assuming a peak Mach-number of unity, together with the Saha equation for ionisation equilibrium. We have included only the inner ``classical'' and saturated regions of the flow. We have neglected those electrons which are expected to be present inside the cloud, as a result of cosmic-ray ionisation, and the refractive index contributions of the neutral gas \citep{D98} and the gravitational field, because all of these contributions depend on the density profile of the neutral material in the cloud and that profile is currently uncertain.

To compute the refraction of radio-waves by this electron density distribution is now a straightforward exercise. It remains to specify the distance of the lens from the observer and the impact parameter of the event; we take these to be $500\;{\rm pc}$ and $1.8$~AU, respectively (notice that this means the line of sight just grazes the edge of the cloud at closest approach). The radio source is assumed to be made up of two components, each of flux 0.3~Jy: an extended region which is so large that it is not subject to any significant magnification, and a compact component which we represent by a circular Gaussian source with a peak brightness temperature of $4\times10^{12}\;{\rm K}$. The resulting model light curves are shown in figure 4. These light-curves have a broad resemblance to those in figure 2, confirming that our physical model is indeed a sensible one.

In detail the model light-curves do not match the data for Q0954+658, but this is only to be expected given the simplistic nature of our lens and source models; we will not attempt a detailed comparison here. Nor do we compare our model with the various ESE lightcurves observed in other sources \citep{F94}; on this point, however, we note that a diverse set  of light-curves can be generated by choosing different values for the various model parameters, so the diversity of behaviour seen in \citet{F94} is not in itself problematic.

\begin{figure}[!ht]
\begin{center}
\includegraphics[width = 12cm, clip = true, bb = 70 350 550 700]{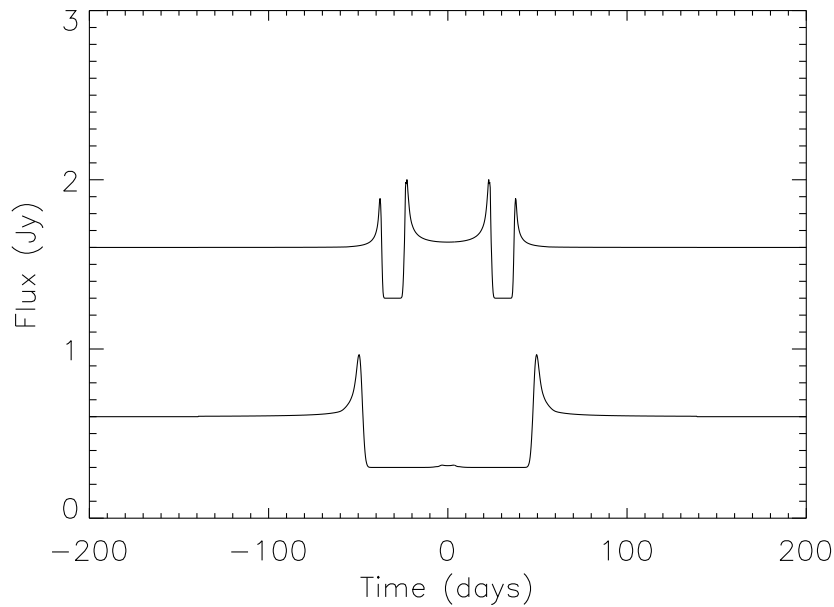}
\end{center}
\vskip -1cm
\caption{Model light-curves -- high and low-frequencies as per figures 1 \&\  2 -- for a compact radio source lensed by an ionised gas flow as described in \S4. In this model a cold, neutral cloud shock-heats the diffuse ISM to a million degrees, and this hot gas evaporates material from the surface of the cloud.}\label{fig4}
\end{figure}

\section{The magnetotail}
The foregoing discussion neglects the interstellar magnetic field. This is a reasonable first approximation for describing the evaporation flow but in the wake of the cloud, where the ionised gas has cooled significantly, the magnetic field is likely to be the dominant contribution to the stress tensor. Here in the magnetotail the field is stretched out along the direction of motion of the cloud. The high conductivity ionised gas is ``frozen'' to the field lines so any inhomogeneities in the gas are fixed laterally but free to equilibrate along the field and should thus be filamentary. Considering that the cloud is moving highly supersonically relative to the ISM, it seems likely that a great deal of inhomogeneity will be generated downstream of the cloud and that the magnetotail will be rich in filamentary structure.  The effect of this material on radio-waves is important: it will introduce large phase changes, and the phase structure will be highly anisotropic with refraction and scattering occurring predominantly perpendicular to the velocity of the cloud (relative to the ISM), projected onto the sky. It is beyond the scope of this paper to attempt to quantify the  structure in the magnetotail, but because the strength and anisotropy of the scattering are both expected to be large  we propose that this region should be considered as a likely site in which IntraDay Variability of quasars \citep{KC97, dB06, B06}, and the parabolic arcs seen in pulsar dynamic spectra \citep{S01, S06} may arise. In connection with these phenomena we note that the transverse velocity of the inhomogeneities is oriented perpendicular to the scattering angle, and does not much affect the ``screen velocities'' inferred from observation as these are sensitive to the velocity component parallel to the scattering angle. Another point is that the magnetotail itself is expected to be much longer than it is wide \citep{SL99}, so if we have correctly identified the physical site of the anomalous scattering phenomena then the scattering screens should have a highly elongated geometry with scattering predominantly perpendicular to the long axis of the screen.  Finally we note that the highly ordered magnetic structure in the magnetotail, including an abrupt field reversal,  should manifest itself in corresponding Rotation Measure structures. 

\section{Summary and conclusions}
Analysis of the data on the ESE in Q0954+658 shows that this radio-wave lensing event is caused by an axisymmetric distribution of ionised gas. The absence of a large number of comparable asymmetric events tells us that the axisymmetry arises from an underlying spherical symmetry. We conclude that each ESE lens is associated with a neutral, self-gravitating gas cloud. The large sky-covering fraction of ESEs robustly indicates that these clouds make an important contribution to the dynamics of the Galaxy, so they are a form of dark matter --- a result which has been noted previously in connection with a flawed physical model \citep{WW98, M00}.

The low levels of rapid variability on the leading edge of the Q0954+658 refraction event, relative to the trailing edge, imply a large difference between the external pressures on these two sides of the cloud. This differential, and the high thermal pressure associated with a spherical lens,  is readily understood as being due to the ram pressure of the diffuse ISM seen in the frame of a cloud moving at hundreds of km/s.  Speeds of this order are unsurprising for a dynamically significant component of the Galaxy. 

Scattering in the magnetotail of the cloud is expected to be strong and highly anisotropic and offers a plausible physical site for two other anomalous scintillation phenomena, namely Intra-Day Variability of quasars and parabolic arcs in pulsar dynamic spectra. 

In summary: although we have little information on ESEs, and our understanding of the phenomenon is currently poor, analysis of the existing data does provide a clear account of the physical environment in which the lenses are formed --- namely cold, neutral, self-gravitating clouds moving at hundreds of km/s through the diffuse ISM. A simple quantitative model of this physical circumstance yields passable light-curves, confirming these model-independent deductions. We conclude that AU-sized gas clouds are an important constituent of the ISM and a major contribution to the mass of the Galaxy.

\acknowledgements 
Many colleagues have contributed their insights on the ESE phenomenon in numerous discussions with the author over a period of several years. In connection with the present paper I would like to acknowledge Mark Wardle, Don Backer, Dave Hollenbach and Chris McKee -- all of whom gave freely of their time to discuss the physics of the specific model which I presented at the SINS conference -- and the SINS conference attendees themselves for their thoughtful comments, which I hope are addressed in this manuscript.

\end{document}